\documentclass[a4paper,11pt]{article}

\usepackage{contribution}

\newcommand{\weblink}[2][]{%
    \ifthenelse{\equal{#1}{}}%
    {\textnormal{\url{#2}}}%
    {\textnormal{\href{#2}{#1}}}%
}

\newcommand{\acknowledgements}[1]{%
  \bigskip\bigskip
  \textsf{\textbf{\Large Acknowledgements}} \\[2ex]
  {#1}
  \bigskip
}

\def\beq{\begin{equation}}
\def\eeq#1{\label{#1}\end{equation}}
\def\eeqn{\end{equation}}

\def\beqa{\begin{eqnarray}}
\def\eeqa#1{\label{#1}\end{eqnarray}}
\def\eeqan{\end{eqnarray}}

\let\bar=\overbar

\def\M{{\cal M}}

\def\Dslash{\not{\hbox{\kern-4pt $D$}}}
\def\dslash{\not{\hbox{\kern-2pt $\del$}}}

\def\msb{{\bar{\ssstyle M \kern -1pt S}}}

\newcommand{\contribution}[7][]{%
  \clearpage
  \thispagestyle{plain}
  \ifthenelse{\equal{#1}{}}
  {\hypersetup{pdftitle={#2}}}
  {\hypersetup{pdftitle={#1}}}
  \hypersetup{pdfauthor={{#3} {#4}}}
  {\centering\normalfont\LARGE\bfseries\sffamily #2 \par\nobreak}
  \lhead{}
  \chead{%
    \textit{\footnotesize XIV International Conference on Hadron Spectroscopy
      (\weblink[\textit{hadron2011}]{http://www.hadron2011.de}), 13-17 June 2011, Munich, Germany}%
  }
  \rhead{}
  \bigskip
  \begin{center}
    {#3} {#4}\ifthenelse{\equal{#6}{}}{}{\footnote{\weblink[#6]{mailto:#6}}}
    \ifthenelse{\equal{#7}{}}{}{#7} \\
    \textit{#5}
  \end{center}
  \bigskip
}

\renewcommand{\abstract}[1]{%
  \begin{center}
    \begin{minipage}{0.85\textwidth}
      \begin{footnotesize}
        #1
      \end{footnotesize}
    \end{minipage}
  \end{center}
  \bigskip
}

\begin{document}

%
%
%
%
%
{  


\renewcommand{\beq}{\begin{equation}}
\renewcommand{\eeq}{\end{equation}}
\renewcommand{\Im}{\text{Im}\,}
\renewcommand{\Re}{\text{Re}\,}
\newcommand{\Mpi}{M_\pi^2}
\renewcommand{\M}{\mathcal{M}}
\newcommand{\meta}{M_\eta}
\newcommand{\mpc}{M_\pi}


%
\contribution
{The role of final-state interactions in Dalitz plot studies}  
{Bastian}{Kubis}  
{Helmholtz-Institut f\"ur Strahlen- und Kernphysik (Theorie) and \\
 Bethe Center for Theoretical Physics, Universit\"at Bonn, Germany}  
{kubis@hiskp.uni-bonn.de}  
{}  
%

%
\abstract{%
Dalitz plot studies for multi-hadron decays of heavy mesons are expected to become 
very important tools for precision investigations of CP violation. 
A thorough understanding of the hadronic final-state interactions is a prerequisite 
to achieve a highly sensitive, model-independent study of CP-violating phases in such processes. 
We illustrate the theoretical tools available, as well as still to be developed, 
from low-to-medium-energy hadron physics for this purpose, and the goals of the informal 
\emph{Les Nabis} network studying these and related problems.
}
%

%
\section{CP-violation in Dalitz plots}

\begin{sloppypar}
A precise study of final-state interactions is increasingly becoming of paramount
importance for our understanding of the most diverse aspects of particle decays
involving hadrons.  
Final-state interactions can be of significance for various reasons: 
if they are strong, they can significantly enhance decay probabilities;
they can significantly \emph{shape} the decay probabilities, most prominently
through the occurrence of resonances; 
besides resonances, also new and non-trivial analytic structures can occur,
such as threshold or cusp effects (for the prominent role cusp effects have played
recently in studying pion--pion interactions, see Ref.~\cite{Cusp} and references therein); 
and finally, they introduce strong or hadronic phases or imaginary parts, the existence of which 
is a prerequisite for the extraction of CP-violating phases in weak decays (see e.g.\ Ref.~\cite{BigiSanda}).
Dalitz plot studies of weak three-body decays of mesons with open heavy flavor (both $D$ and $B$)
are expected to acquire a key role in future precision investigations of CP violation,
due to their much richer kinematic freedom compared to the (effective) two-body 
final states predominantly used to study CP violation at the $B$ factories.  
In many cases, the branching fractions are significantly larger; furthermore,
the resonance-rich environment of multi-meson final states may help to enlarge small
CP phases in parts of the Dalitz plot, and differential observables may allow 
to obtain information on the operator structure that
drives CP violation beyond the Standard Model, once it is observed.
Since the results from the $B$ factories have shown that the Cabibbo--Kobayashi--Maskawa theory~\cite{CKM}
represents at least the dominant source of CP violation, our long-term goal will be to find other sources 
of CP violation that contribute additional, smaller effects. 
For this purpose, clearly extremely accurate measurements \emph{and} means of theoretical interpretation
are required.
Strong evidence for CP violation in three-body final states has already been reported
for $B^\pm \to K^\pm \pi^\mp \pi^\pm$~\cite{Belle_Babar-BCP}, e.g.\ with a $3.7\sigma$ signal
in the effective $K\rho$ channel, while only negative results exist 
for $D$ decays so far~\cite{Babar-DCP}.
\end{sloppypar}

There are different possibilities how to analyze CP violation in Dalitz plots.  
One suggestion is a strictly model-independent extraction from the data 
directly~\cite{susanmodelindep,miranda}, e.g.\ using the \emph{significance}~\cite{miranda}
variable defined as
\beq
{}^{\rm Dp}S_{\rm CP}(i) \doteq \dfrac{N(i)-\bar N(i)}{\sqrt{N(i)+\bar N(i)}} ~,\eeq
where $N(i)$ and $\bar N(i)$ denote the event numbers of CP-conjugate decay modes
in a specific Dalitz plot bin $i$.
CP violation can then be identified in a deviation from a purely Gaussian distribution
in the significance plots.  The significance method allows to study \emph{local} asymmetries
and requires no theoretical input at all.

An alternative approach, in contrast, makes use of information on the strong amplitudes as input.
To see why this may be advantageous, we consider the following toy model:\footnote{I am grateful to 
C.~Hanhart for providing me with this example.}  consider event number distributions 
given by a (Breit--Wigner) resonance signal (of mass $M_{\rm res}$ and width $\Gamma_{\rm res}$) on a certain 
\begin{figure}[tb]
  \begin{center}
    \includegraphics[width=0.7\textwidth,clip]{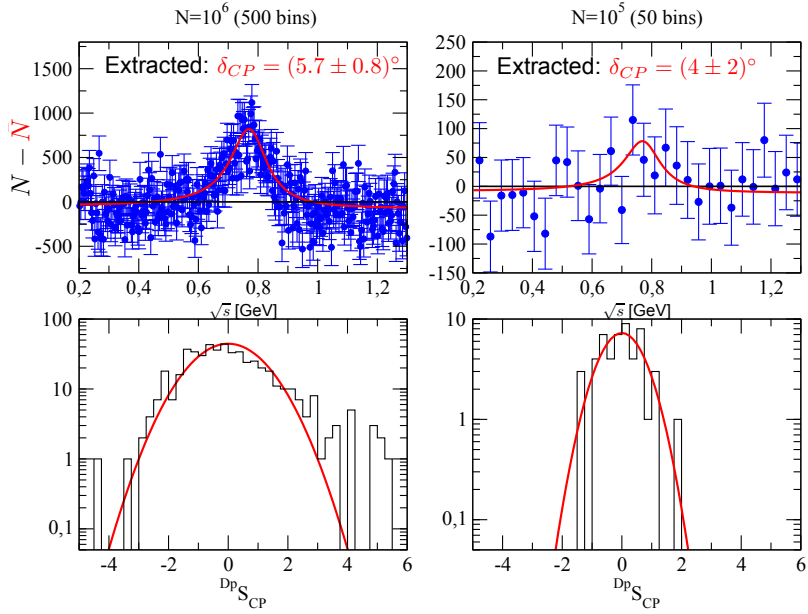} 
    \caption{Toy example for extraction of a CP-violating phase, for high- (left) and low- (right) statistics 
    samples, using the known Breit--Wigner shape (top) versus the significance variable (bottom). 
    Input parameters used are $\delta_{\rm CP} = 5^\circ$, $M_{\rm res} = 0.77\,\text{GeV}$, 
    $\Gamma_{\rm res} = 0.15\,\text{GeV}$. Figure courtesy of C.~Hanhart.}
    \label{fig:demo_BW}
  \end{center}
\end{figure}
background, with a CP-violating phase $\delta_{\rm CP}$, according to
\beq
N, \, \bar N = \alpha + \beta \,\Re \Big\{\frac{e^{\pm i \delta_{\rm CP}}}{s-M_{\rm res}^2+i M_{\rm res} \Gamma_{\rm res}}\Big\} \quad\Rightarrow\quad
N-\bar N = 
\frac{\sin\delta_{\rm CP} \times 2\beta M_{\rm res}\Gamma_{\rm res}}{(s-M_{\rm res}^2)^2+(M_{\rm res} \Gamma_{\rm res})^2} ~. \label{eq:demo_BW}
\eeq
Figure~\ref{fig:demo_BW} shows the count-rate difference and the corresponding significance distributions
for two simulated cases of pseudo-data of different statistical weight: 
while in the high-statistics case, there indeed seems 
to be a deviation from Gaussian distribution in the significance, this is definitely lost in the case
of less events.  In contrast, fitting the data with the functional form~\eqref{eq:demo_BW} still allows 
to extract $\delta_{\rm CP}$ with some (limited) accuracy even in the sparser sample.
So while the theoretical assumptions and prejudices going into such an analysis clearly have to be
very carefully judged, their benefit in terms of vastly increased sensitivity is also obvious.

In the following, we will therefore briefly sketch some of the tools available
to analyze the hadronic amplitudes of the (light) final-state particles (such as pions and kaons).

\section{Pion--pion scattering: Roy equations}

Analyticity, unitarity, and crossing symmetry provide a high degree of constraint
for the pion--pion scattering amplitude, which can be exploited using dispersion relations.
Starting from a twice-subtracted dispersion relation at fixed Mandelstam variable $t$,
\beq
T(s,t) = c(t) + \frac{1}{\pi} \int_{4\Mpi}^\infty ds' \Big\{
\frac{s^2}{{s'}^2(s'-s)} + \frac{u^2}{{s'}^2(s'-u)} \Big\} \Im T(s',t) ~,
\eeq
the subtraction function $c(t)$ can be determined from crossing symmetry.
Projecting onto partial waves $t_J^I$ of definite angular momentum $J$ and isospin $I$, 
one obtains a coupled system of partial-wave integral equations,
\beq
t_J^I(s) = k_J^I(s) + \sum_{I'=0}^2 \sum_{J'=0}^\infty  \int_{4\Mpi}^\infty ds'
K_{JJ'}^{II'}(s,s') \Im t_{J'}^{I'}(s') ~, \label{eq:partialwave}
\eeq
where the kernels $K_{JJ'}^{II'}(s,s')$ are kinematical functions known analytically. 
The subtraction polynomial $k_J^I(s)$ contains the $\pi\pi$ scattering lengths as the
only free parameters; these may in turn be further constrained by matching to chiral
perturbation theory~\cite{CGL}.   Equation~\eqref{eq:partialwave} can finally be turned into 
a coupled set of equations for the \emph{phase shifts}, the Roy equations~\cite{Roy}, 
by assuming elastic unitarity in the form
\beq
t_J^I(s) = \frac{e^{2i\delta_J^I(s)}-1}{2i\sigma} ~, \quad
\Im t_J^I(s) = \sigma |t_J^I(s)|^2 ~, \quad
\sigma = \sqrt{1-\frac{4\Mpi}{s}} ~.
\eeq
Modern precision analyses of the Roy equations have been performed in Ref.~\cite{ACGL+GarciaMartin},
and similarly for pion--kaon scattering~\cite{Buettiker}; compare also the discussion 
of $\gamma\gamma\to\pi\pi$ at this conference~\cite{ggpipi}.
These provide us with high-precision parameterizations of the most relevant scattering amplitudes 
for light mesons appearing in the final states of heavy-meson decays.

\section{Analyticity and unitarity for form factors}

\begin{figure}[tb]
  \begin{center}
    \includegraphics[width=0.55\textwidth]{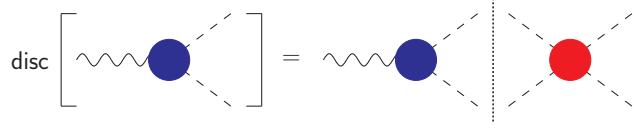} 
    \caption{Graphical representation of the consequence of analyticity and unitarity for form factors.}
    \label{fig:FFunit}
  \end{center}
\end{figure}

Final-state interactions between \emph{only two} strongly interacting particles
can be described in terms of form factors, which in turn can be linked to the properties
of scattering amplitudes using analyticity and unitarity.  As illustrated in Fig.~\ref{fig:FFunit},
the unitarity relation for a form factor $F_J^I(s)$ (here: of the pion) reads
\beq
\Im F_J^I(s) = F_J^I(s) \times \theta \big(s-4\Mpi\big) \times \sin\delta_J^I(s) e^{-i\delta_J^I(s)} ~,
\label{eq:FF}
\eeq
from which one immediately deduces Watson's final-state theorem~\cite{Watson}: the form factor
shares the phase $\delta_J^I(s)$ of the (elastic) scattering amplitude.
The solution to Eq.~\eqref{eq:FF} is obtained in terms of the Omn\`es function~\cite{Omnes},
\beq
F_J^I(s) = P_J^I(s) \Omega_J^I(s) ~, \quad
\Omega_J^I(s)=\exp\bigg\{\frac{s}{\pi}\int\limits_{4\Mpi}^\infty ds'\frac{\delta_J^I(s')}{s'(s'-s)} \bigg\} ~,
\label{eq:Omnes}
\eeq
where $P_J^I(s)$ is a polynomial. Note that the Omn\`es function is completely given in terms of the phase shift.
A classic application of such a form factor representation is the pion vector form factor $F_V^\pi(s)$, 
which rather than by Eq.~\eqref{eq:Omnes} is written in a more refined way as
\beq
F_V^\pi(s) = \Omega_1^1(s) G_\omega(s) \Omega_{\rm inel}(s) ~,
\eeq
where $G_\omega(s)$ takes into account $\rho-\omega$ mixing, 
and $\Omega_{\rm inel}(s)$ parameterizes inelasticities, effectively setting in above $s \gtrsim (M_\pi+M_\omega)^2$,
using conformal mapping techniques~\cite{Troconiz}.
Such form factor representations can be used for analyses of the $e^+e^-\to\pi^+\pi^-$ data
to reduce the error of the hadronic contribution to the muon $g-2$, or to check the 
compatibility of the data with analyticity and unitarity~\cite{Bern:piFF}.
Note that the effects of chiral dynamics are particularly important for \emph{scalar} form 
factors, where a parameterization in terms of Breit--Wigner resonances 
can lead to completely wrong phase motions (see e.g.\ Ref.~\cite{susanandulf} for the context
of $B\to 3\pi$ decays).

\section{Dispersion relations for three-body decays}

\begin{figure}[tb]
  \begin{center}
    \includegraphics[width=0.25\textwidth]{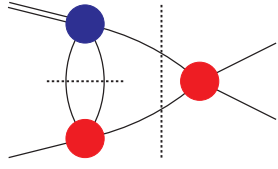} 
    \caption{Example for the complication of the analytic structure of 4-point functions 
             through crossed-channel effects, here for the decay of an $\eta$ (double line)
             into three pions (single lines).}
    \label{fig:4Punit}
  \end{center}
\end{figure}

\begin{sloppypar}
The application of dispersion relations to three-body decays is more complicated than 
the treatment of form factors due to the more involved analytic structure, and the possibility 
of crossed-channel rescattering (compare also Ref.~\cite{Magalhaes} reported at this conference); 
see Fig.~\ref{fig:4Punit} for a depiction of the complication
of the unitarity relation.  We discuss here the (low-energy) example of $\eta\to 3\pi$ decays,
which has received much renewed attention recently~\cite{Lanz,SDK+Kampf} due to its importance 
for the extraction of the light quark mass ratios.
One starts by decomposing the amplitude $\M(s,t,u)\propto\mathcal{A}(\eta\to\pi^+\pi^-\pi^0)$ 
into partial waves of isospin $I$ according to~\cite{Stern:pipi,AnisovichLeutwyler}
\beq
\M(s,t,u) = \M_0(s) + (s-t)\M_1(u) + (s-u)\M_1(t)+ \M_2(t) + \M_2(u) - \frac{2}{3} \M_2(s) ~,
\label{eq:etaPW}
\eeq
where the $\M_I(s)$ are functions of one variable only, with only a right-hand cut.
Equation~\eqref{eq:etaPW} is exact as long as discontinuities of D- and higher partial waves are neglected.
The unitarity relation for the $\M_I(s)$, 
\beq
\Im \M_I(s) = \big\{\M_I(s) + \hat\M_I(s)\big\} \times \theta \big(s-4\Mpi\big) \times \sin\delta_I(s) e^{-i\delta_I(s)} ~
\label{eq:ImM}
\eeq
(we now ignore the angular-momentum indices), is then complicated compared to Eq.~\eqref{eq:FF}
by \emph{inhomogeneities} $\hat\M_I(s)$, which are given by angular averages over the $\M_I$ according
to
\begin{align}
\hat\M_0(s) & =\frac{2}{3}\langle\M_0\rangle(s)
+\frac{20}{9}\langle\M_2\rangle(s)+2(s-s_0)\langle\M_1\rangle(s)
+\frac{2}{3}\kappa(s)\langle z\M_1\rangle(s)  ~, \nonumber\\
\langle z^nf\rangle(s) &=\frac{1}{2}\int_{-1}^1 dz\,z^n f\big(\tfrac{1}{2}(3s_0-s+z\kappa(s))\big) ~, 
\quad s_0 = \frac{1}{3}\big(\meta^2+3\mpc^2\big) ~,\nonumber\\[-1mm]
\kappa(s)&=\sqrt{(s-(\meta+\mpc)^2)(s-(\meta-\mpc)^2)}\times\sqrt{1-\frac{4\,\Mpi}{s}} ~,\label{eq:Mhat}
\end{align}
and similarly for the other $\hat\M_I$.
Note that the angular integration including the $\kappa(s)$ function is non-trivial and generates
a complex analytic structure, including three-particle cuts due to the fact that the $\eta$ is 
unstable and decays.
The analog to the Omn\`es solution~\eqref{eq:Omnes} are then integral equations involving
the inhomogeneities~\cite{AnisovichLeutwyler}
\beq
\M_0(s) = \Omega_0(s)\biggl\{\alpha_0+\beta_0\,s+\gamma_0\,s^2+\frac{s^3}{\pi}\int_{4\Mpi}^\infty\frac{ds'}{{s'}^3}\frac{\sin\delta_0(s')\hat\M_0(s')}{|\Omega_0(s')|(s'-s-i\epsilon)}\biggr\} ~, \label{eq:inhomOmnes}
\eeq
with subtraction constants $\alpha_0,\,\beta_0,\,\gamma_0$, and similar forms 
for the other partial waves.  (See Ref.~\cite{KhuriTreiman+Aitchison} for earlier, related
formulations.) 
\begin{figure}[tb]
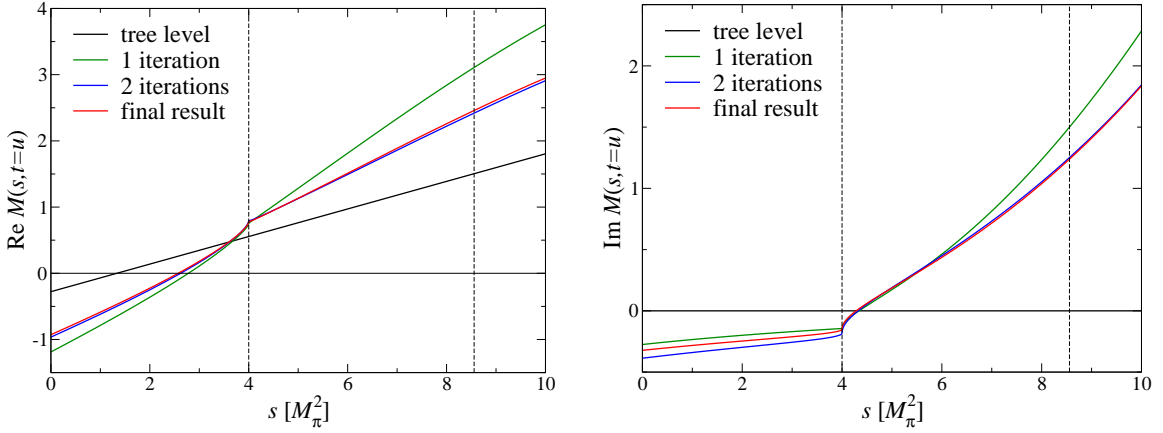

  \begin{center}
    \includegraphics[width=0.48\textwidth]{ReAmp} \hfill
    \includegraphics[width=0.48\textwidth]{ImAmp}
    \caption{Real and imaginary part of the $\eta\to\pi^+\pi^-\pi^0$ amplitude $\M(s,t=u)$. 
    Compare also Ref.~\cite{Lanz}.}
    \label{fig:eta3pi}
  \end{center}
\end{figure}
Equations~\eqref{eq:Mhat} and \eqref{eq:inhomOmnes} can then be solved iteratively,
e.g.\ matching the subtraction constants to chiral perturbation theory, see Fig.~\ref{fig:eta3pi}.
The iteration converges fast, with the second iteration already very close to the final result.
\end{sloppypar}

The method sketched here is currently also applied to other light-meson decays such as 
$\eta'\to\eta\pi\pi$ or $\omega,\,\phi\to 3\pi$~\cite{inprogress}. Challenges to be faced when 
extending this formalism to heavy-meson decays include the necessity to treat systems of
integral equations when coupled channels within one partial wave cannot be ignored, or
inelasticities are not negligible.  
In particular when considering $B$-meson decays, elastic unitarity will surely not be sufficient.
It will have to be checked when a purely perturbative
treatment of crossed-channel effects is feasible (compare e.g.\ Ref.~\cite{Liu}), and when 
higher partial waves become important.
To investigate these and related questions is part of the program of the informal
{\it Les Nabis} network~\cite{LesNabis}, which 
brings together physicists from theory and experiment in heavy- and light-quark physics and
aims at optimizing future Dalitz plot studies along the lines sketched here---with the 
strong goal to better interpret the mechanism of CP violation in nature, yet at the same time
teaching us important lessons on nonperturbative strong interactions.

\newpage

\acknowledgements{%
I would like to thank the organizers of Hadron 2011 for inviting me to such an inspiring conference,
C.~Hanhart for stimulating discussions concerning the material presented here,
and I.~I.~Bigi as well as M.~Hoferichter for useful comments on this write-up.
Partial financial support 
by DFG (SFB/TR 16, ``Subnuclear Structure of Matter''),
by the project ``Study of Strongly Interacting Matter'' 
(HadronPhysics2, grant 227431) under the 7th Framework Program of the EU,
and by the Helmholtz Association providing funds to the virtual 
institute ``Spin and strong QCD'' (VH-VI-231) is gratefully acknowledged.
}

%

}  


\end{document}